# Gravitational Lensing of Quasars by Spiral Galaxies

Abraham Loeb

*Harvard Astronomy Department, 60 Garden Street, Cambridge, MA 02138*

Gravitational lensing by a spiral galaxy occurs when the line-of-sight to a background quasar passes within a few kpc from the center of the galactic disk. Since galactic disks are rich in neutral hydrogen (cf. Fig. 1), the quasar spectrum will likely be marked by a damped Ly$\alpha$ absorption trough at the lens redshift. Therefore, the efficiency of blind searches for gravitational lensing with sub-arcsecond splitting can be enhanced by 1–2 orders of magnitude [1] by selecting a subset of all bright quasars which show a low-redshift ($z \lesssim 1$) damped Ly$\alpha$ absorption with a high HI column density, $N \gtrsim 10^{21}$ cm$^{-2}$.

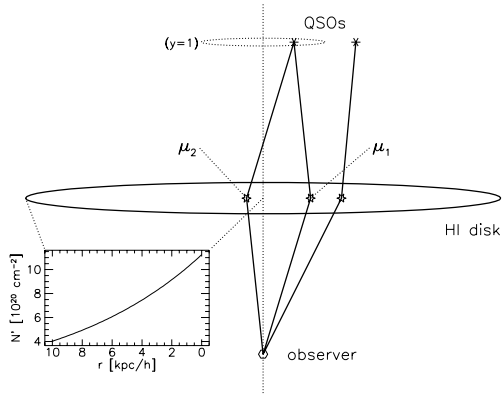

**Figure 1.** Lensing geometry of a spiral galaxy. Quasars are multiply imaged if their projection lies within the dotted circle. The magnification factors of the two images of the central quasar are $\mu_1$ and $\mu_2$. The insert displays the characteristic HI column density as a function of radius on the face of the HI disk of the lens.

The magnification bias due to lensing changes the statistics of damped Ly$\alpha$ absorbers (DLAs) in quasar spectra by bringing into view quasars that are otherwise below the detection threshold. For optical observations this effect is often counteracted by dust extinction in the lensing galaxy. The combination of lensing and dust extinction results in a net distortion of the intersection probability of HI column densities for spirals.[2] The distortion shows a peak which rises above the unlensed probability distribution. The peak disappears at high column densities because lensing bends the light rays and prevents the brightest quasar image from crossing the HI disk at an arbitrarily small impact parameter. The width of this peak is ultimately determined by the variance of HI profiles and potential wells in the absorber population.[3]

Spiral lenses are difficult to find because their characteristic image sepa-ration is a fraction of an arcsecond. The multiple image signature of lensing could, however, be identified spectroscopically and without a need for high-resolution imaging. In the case of the double image system illustrated in Figure 1, the HI absorption spectrum of the quasar would show a generic double–step profile due to the superposition of the two absorption troughs of the different images. This profile is shown schematically in Figure 2. The different images cross the absorbing disk at different impact parameters and therefore have different damped Ly$\alpha$ widths, $W_i \propto N_i^{1/2}$, $(i = 1, 2)$. The metal absorption lines would probe different kinematic components in the two images. In the absence of extinction by dust, the depth of a given step is fixed by the magnification of the corresponding image, $\mu_i$, which reflects the fraction of all detected photons that probe the column density associated with that image. This method for identifying lenses might be contaminated by intrinsic fluctuations in realistic absorption troughs. As a first step towards establishing the feasibility of this technique, one might measure the absorption spectrum of quasars which *are known* to be lensed by a spiral galaxy. The spiral lens B0218+357 constitutes a generic example for such a case. The lensed quasar was discovered in the radio and found to be an Einstein ring of radius $\sim 0.3''$ with two compact components.[4, 5] Observations of 21 cm absorption in this lens[6] indicate a high HI column density, $N = 4 \times 10^{21}$ cm$^{-2}$ $(T_s/100 \text{ K})/(f/0.1)$, where $T_s$ is the spin temperature of the gas and $f$ is the HI covering factor. The structure of the Ly$\alpha$ absorption trough in this lens will be measured spectroscopically in the near future with the Hubble Space Telescope[7]. Systems similar to B 0218+357 should be common among quasars with strong damped Ly$\alpha$ absorption.

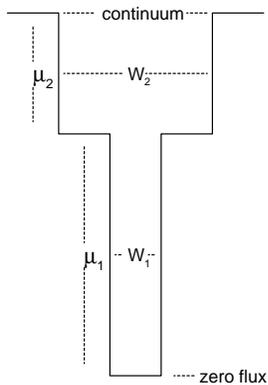

**Figure 2.** Structure of the Ly$\alpha$ absorption spectrum of a quasar lensed by a spiral galaxy. The double trough profile allows *spectroscopic* identification of a gravitational lens without the need for sub–arcsecond imaging.

Recent microlensing searches [8] indicate that a non–negligible fraction of the halo mass of the Milky–Way galaxy might be in the form of Massive Compact Halo Objects (MACHOs). As a supplement to local searches, it would be interesting to examine whether MACHOs populate the halos of high–redshift spirals. While the microlensing probability is only $\sim 10^{-6}$ in the Milky-Way halo ($\sim 10$ kpc), its value increases up to unity in the cores of halos at cosmological distances ($\sim 5$ Gpc). This follows from the linear dependence of the lensing cross–section on the observer–lens distance for a source at infinity. In particular, one could predict the expected microlensing probability (due to disk stars or MACHOs) in distant spiral galaxies which show up as DLAs in quasar spectra [9]. The obvious advantage of DLAs is that they are selected based on their proximity on the sky to a quasar.

The characteristic Einstein radius of a solar mass lens at a cosmological distance is $\sim 5 \times 10^{16}$ cm, comfortably in between the scales of the continuum–emitting accretion disk ($\lesssim 10^{15}$ cm) and the broad line region ($\sim 3 \times 10^{17}$ cm) of a bright quasar. This implies that a microlensing event would magnify the continuum but not the broad lines emitted by the quasar. As a result, the equivalent width distribution of broad lines [10] will be systematically shifted towards low values in a sample of microlensed quasars. [11, 12] The microlensing probability of a spiral lens is strongly enhanced if the halo of the galaxy is made of MACHOs. [9] In this case, the distortion imprinted by microlensing on the equivalent width distribution of quasar emission lines could be detected in a relatively small sample of only $\sim 10$ DLAs with HI column densities $N \gtrsim 10^{21}$ cm$^{-2}$ and absorption redshifts $z_{\rm abs} \lesssim 1$. [9] In addition, about a tenth of all quasars with DLAs ($N \gtrsim 10^{20}$ cm$^{-2}$) might show excess variability on timescales of order 1–10 years. [9] A search for these signals would complement microlensing searches in local galaxies and calibrate the MACHO mass fraction in galactic halos at high redshifts.

Finally, we note that a highly inclined disk could contribute substantially to the lensing cross–section of spiral galaxies. In particular, a quasar behind a razor–thin edge–on disk with a projected mass-per-unit-length $\mu$, would always acquire multiple images with separation $\sim G\mu/c^2$, irrespective of its radial distance from the lens center. Even after averaging over all disk inclinations, the cross-section for lensing by spirals could still be substantially larger than the standard singular-isothermal-sphere value. [13] The lens system B 1600+434 was recently reported [14] to have an edge-on disk in between the lensed quasar images. It therefore provides an excellent candidate for a disk–lensing configuration. Since extinction by dust is common in spiral lenses such as B 0218+357 or B 1600+434, the incidence of spiral lenses could be more common in radio surveys such as CLASS [15].

**Acknowledgments**

I thank Matthias Bartelmann and Rosalba Perna for many discussions on the subjects mentioned in this contribution.